\documentclass[12pt]{article} 

\usepackage{epsfig}

\def\k40{\rm{^{40}K}}
\def\cu63{\rm{^{63}Cu}}
\def\kstable{\rm{^{39}K}}

\begin{document}
\hspace*{1cm} \hfill \today

\begin{center}
{\Large  \bf
Measurement of
Ultra-Low
Potassium Contaminations
with Accelerator Mass Spectrometry
}\\
\vskip 0.5cm
\large
K.J.~Dong$^{a,b}$,
H.T.~Wong$^{b,}$\footnote{Corresponding~author:
Email:~htwong@phys.sinica.edu.tw;
Tel:+886-2-2789-6789;
FAX:+886-2-2788-9828.},
M.~He$^{a}$,
S.~Jiang$^{a}$,\\
J.Z.~Qiu$^{a,c}$,
Y.J.~Guan$^{a,d}$,
S.Y.~Wu$^{a}$,
J.~Yuan$^{a}$

\normalsize
\vskip .2cm

\begin{flushleft}
{$^{a}$\rm 
Department of Nuclear Physics,
China Institute of Atomic Energy, Beijing 102413.\\}
{$^{b}$\rm
Institute of Physics, Academia Sinica, Taipei 11529.\\}
{$^{c}$\rm
Armed Police Force Academy, Langfang 065000.\\}
{$^{d}$\rm
Department of Physics, Guanxi University, Nanning 530004.\\}
\end{flushleft}

\end{center}
\vskip 0.5cm
\begin{abstract}

Levels of trace radiopurity in active
detector materials is a subject of major concern
in low-background experiments. 
Among the radio-isotopes,
$\k40$ is one of the most abundant and yet whose
signatures are difficult to reject.
Procedures were devised to measure
trace potassium
concentrations in the inorganic salt
CsI as well as in organic liquid scintillator (LS)
with Accelerator Mass Spectrometry (AMS),
giving, respectively, the 
$\k40$-contamination levels 
of $\sim  10^{-10}$ and $\sim 10^{-13}$~g/g.
Measurement flexibilities and
sensitivities are improved over 
conventional methods.
The projected limiting sensitivities 
if no excess of potassium signals
had been observed over background
are $8 \times 10^{-13}$~g/g and $3 \times 10^{-17}$~g/g
for the CsI and LS, respectively.
Studies of the LS samples indicate that the
radioactive contaminations come mainly in
the dye solutes, while the base solvents are
orders of magnitude cleaner.
The work demonstrate the
possibilities of measuring naturally-occurring
isotopes with the AMS techniques.

\end{abstract}

\begin{flushleft}
\end{flushleft}

\vfill

\newpage

\section{Introduction}

Accelerator Mass Spectrometry (AMS)~\cite{amsoverview}
is an established technique for the detection
and measurement of trace level of 
long-lived radionuclides~\cite{amsradio}
such as $^{14}$C, $^{10}$Be, $^{36}$Cl.
It improves over conventional techniques
such as  neutron activation analysis (NAA),
inductively-coupled plasma mass spectrometry (ICP-MS)
as well as 
secondary ion mass spectrometry (SIMS),
and has been successfully applied 
to a wide variety of disciplines like 
archeology, geology, environmental science, 
biomedicine, safeguards of nuclear
materials and so on~\cite{amsapplications}.

Measurement of trace
concentrations of naturally-occurring,
as well as cosmic-ray or fission induced
radioactive isotopes
is an important technique with
major impact to low background experiments~\cite{pdg06}.
The TEXONO Collaboration is pursuing
a research program in low energy
neutrino and astroparticle physics~\cite{texono}.
An important aspect is to devise 
procedures for measuring such trace radiopurity using
the AMS techniques.
Significant
improvements in sensitivities and flexibilities
over existing methods can be expected.
Results were published already for the
measurement of trace $^{129}$I~\cite{ciaei129}
in inorganic crystal and organic liquid scintillator (LS).
The isotope $^{129}$I is
a long-lived fission fragment, 
where the corresponding AMS procedures 
are already matured and well-established. 
In this article.
we report on our efforts and results 
on potassium ($\kstable$).
The isotope $\kstable$ is stable and
exists in large abundance in nature 
$-$ and there are
no prior attempts to 
measure trace abundance of it with AMS. 
These poses new experimental challenges 
and the results reported here represent
advances in the field.

Potassium has three naturally-occurring isotopes:
$\kstable$, $\k40$ and $^{41}$K with 
isotopic abundances of 93.26\%, 0.0117\% and 6.73\%, 
respectively. Among these, $\kstable$ and $^{41}$K
are stable while $\k40$ has 
a half-live of $1.248 \times 10^9$~years, where the
dominant decay channels are:
\begin{equation}
\rm{
\k40 ~ \rightarrow ~ ^{40}Ca ~ + ~ e^- ~ + ~ \bar{\nu _e} 
~~~~  ( ~ BR=89\% ~ ; ~ Q_{\beta ^-}=1.31~MeV ~ ) ~, 
}
\end{equation}
and
\begin{eqnarray}
\k40 ~ +  ~ e^-  & ~ \rightarrow ~ & \rm{ ^{40}Ar^*  + \nu _e ~~~~
( ~  BR=11\% ~ ; ~ Q_{EC}=1.51~MeV ~ ) ~ ,} \\
\rm{ ^{40}Ar^* }  & ~ \rightarrow ~ & 
\rm{ ^{40}Ar ~ + ~ \gamma  ~~~~
( ~ E_{\gamma}=1.461~MeV  ~ ) ~. }
\end{eqnarray}
Both decays pose severe background problems to
low-count-rate experiments where the signals are
at the sub-MeV energy range, such as 
the studies of pp and $^7$Be solar neutrinos,
the searches of Cold Dark Matter, as well as of
neutrino magnetic moments with reactor neutrinos~\cite{pdg06}. 
Potassium contaminations in the active detector materials
will give rise to a continuous $\beta$-decay background
up to 1.31~MeV, while those in the passive materials
in the vicinity of the detector will produce
$\gamma$-rays at 1.46~MeV which, after Compton
scatterings, can lead to background at the sub-MeV
range. The suppression and an efficient measurement
technique of ultra-low amount of potassium in a wide
spectrum of materials are therefore an issue of
crucial significance.
As an illustrative example, 
in order for
the BOREXINO and KamLAND experiments 
to achieve the goals of studying the low energy
$^7$Be solar neutrinos with LS,
the required contaminations of $\k40$ should be
less than $10^{-18}$~g/g.  

While the radioactive isotope of interest
to low background experiments is $\k40$,
the AMS techniques devised in this work are
for the measurement of $\kstable$.
The $\kstable$-to-$\k40$ ratio of 7971:1 is constant
for naturally-occurring materials.
Therefore, a measurement of ultra-low
$\k40$ radio-purities is equivalent to one on the level
of $\kstable$-contaminations, but with a
sensitivity enhancement factor of almost $10^{4}$.
In addition, $\k40$ measurements by the
AMS techniques are vulnerable to
background contaminations by $^{40}$Ca, a stable isotope
which exists in abundance in nature.

The sensitivities of the various techniques to the
measurement of ultra-low $\k40$ contaminations
are shown in Table~\ref{compare}.
Photon counting of the 1.461~MeV $\gamma$-line 
by high-purity germanium detectors
(HPGe)~\cite{hpge} is limited by the ambient background.
The Counting Test Facility (CTF)~\cite{ctf} at the
Gran Sasso Laboratory is a 4~ton LS
detector. It took $\sim$30~days to
measure only a single sample of LS, and
is therefore of limited flexibilities.
To address these drawbacks, 
the NAA techniques for potassium~\cite{munich,alabama} 
were recently developed, where 
the sensitivities of $10^{-15} - 10^{-16} ~ {\rm g/g}$
were achieved. In this paper,
we present results on a complementary approach 
using the AMS techniques.

\begin{table}
\begin{center}
\begin{tabular}{|l|ccccc|}
\hline
Techniques & HPGe & CTF & \multicolumn{2}{c}{NAA} & AMS \\
Reference & \cite{hpge} & \cite{ctf}
& \cite{munich} & \cite{alabama} & This Work \\   \hline \hline
Sample Size & $\sim$1~kg & 4~ton & 202~g & 353~g & 86.5~g \\
Sample Type & No & Only LS & No & No & No \\ \hline
$[ \k40 / {\rm sample} ]$ (g/g) 
& $< 10^{-10} ~ ^{\ast}$ & 
$< 10^{-15} ~ ^{\dagger} $ & 
$< 4 \times 10^{-16} ~ ^{\dagger} $ & 
$< 2 \times 10^{-15} ~ ^{\dagger} $ 
& $\sim 10^{-13} ~ {\rm (LS)} ^{\ddagger} $  \\
& & & & & $\sim 10^{-15} ~ {\rm (PC)} ^{\ddagger} $  \\
& & & & & $< 3 \times 10^{-17} ~ ^{\amalg} $  \\ \hline
\end{tabular}
\end{center}
$^{\ast}$ Typical sensitivities for a variety of samples \\ 
$^{\dagger}$ Custom-prepared ultra-pure liquid scintillator samples\\
$^{\ddagger}$ Commercially produced liquid scintillator (LS) and
pseudocumene (PC) samples\\
$^{\amalg}$ Projected limit if no $\kstable$ signals are observed
\caption{
Comparisons of the
various techniques for ultra-low potassium measurements.
}
\label{compare}
\end{table}

The AMS techniques were adopted to tackle this
problem because of its promises on sensitivities
and versatilities.
An organic liquid and an inorganic
salt were selected for studies. 
Once the procedures are established, 
different batches of samples, each requiring little mass, 
can be measured in relatively short time.
The techniques developed can be easily
adapted to other isotopes in trace concentrations,
and with different carrier materials.

In a broader context, the studies represent
also the direction of extending the 
frontiers of the AMS techniques
to include stable and naturally-occurring isotopes,
complementary to the research efforts
by other groups working
on precious metals in minerals~\cite{amsmetal} as well as 
various impurities in semiconductors~\cite{amssi}.
The AMS detection limits for
stable isotopes are in general of the range of $10^{-12}$,
due to the difficulties of establishing an 
ultra-clean condition in an accelerator environment. 
The techniques of background suppression 
are being intensely pursued by the 
communities working on low-background
experiments. Future advances in these domains
may also lead to improvement in the AMS sensitivities
on the naturally-occurring isotopes.

\section{Experimental Set-Up and Procedures}

\begin{figure}
\centerline{
\epsfig{file=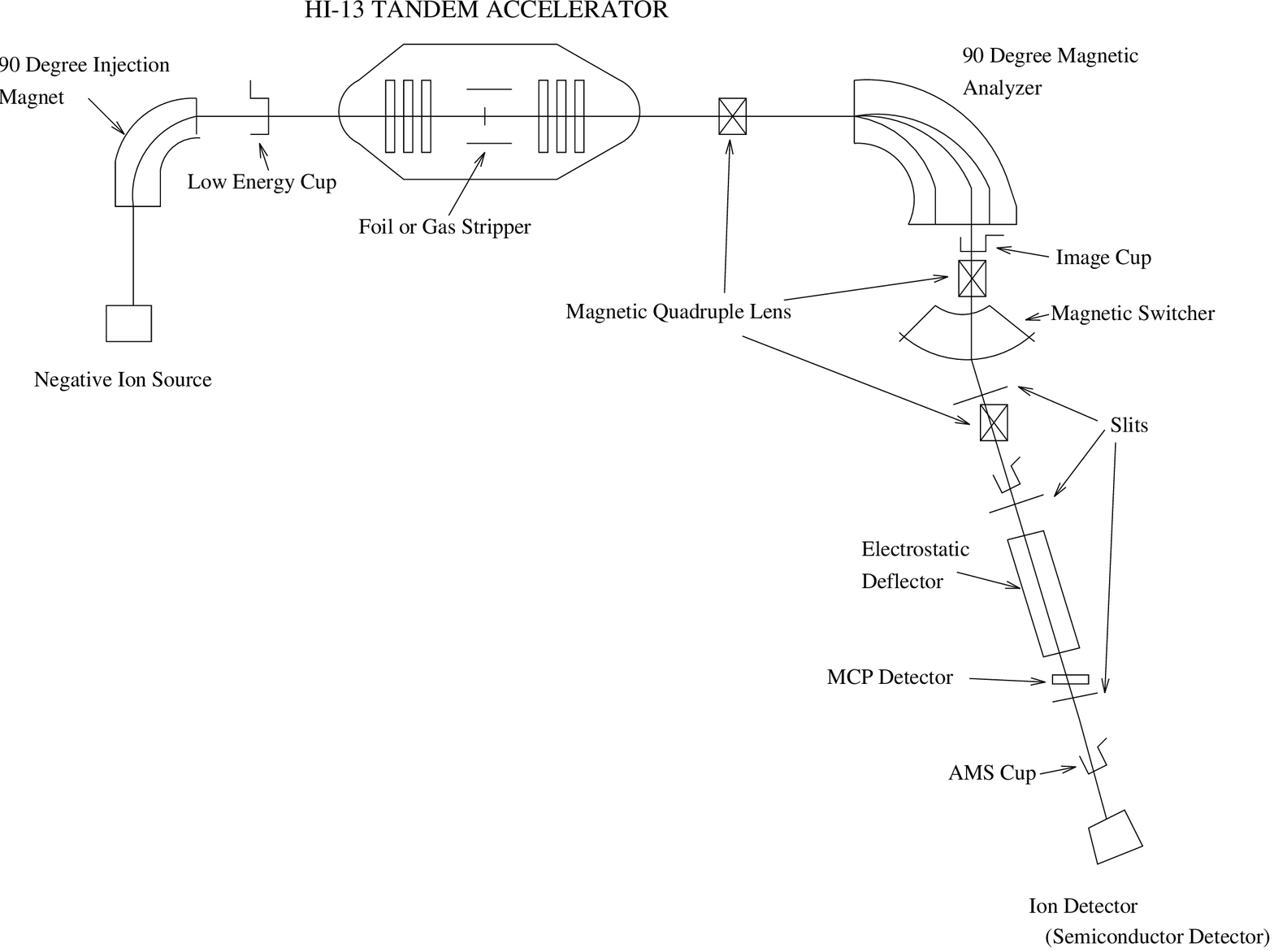,width=12cm,angle=270}
}
\caption{
Schematic layout of the 
13~MV Tandem Accelerator Facility at CIAE
when it is optimized for AMS applications.
}
\label{ciaeams}
\end{figure}

The investigations reported in this
article were conducted at 
at the 13~MV Tandem Accelerator Facility
China Institute of Atomic Energy (CIAE)~\cite{ciaeamsref}.
The optimized configuration for AMS applications
is shown schematically in Figure~\ref{ciaeams}.

An organic liquid and an inorganic
salt were selected as the sampling
materials since
they require different experimental procedures
and systematic effect considerations.
The organic liquid studied
is the standard 
liquid scintillator (LS) mixture $-$
the dye PPO [2,5-diphenyloxazole]
in powder form
dissolving in the organic solvent pseudocumene (PC)
[1,3,5-trimethylbenzene]\footnote{Supplier: Gaonengkedi 
Science \& Technology Co. Ltd., China}.
The inorganic salt selected
was CsI powder\footnote{Supplier: Chemtall GMBH, Germany},
since CsI(Tl)
crystal scintillators are being used in
reactor neutrino~\cite{texonocsi}
and dark matter~\cite{kims} experiments.
The processing and measurement procedures
are expected to be similar 
with other organic solvent and dyes
or inorganic salts.

\subsection{Pre-processing}
\label{sect::process}

No chemical procedures are necessary for the
CsI powder which was directly used in the
AMS measurement. However, CsI is a
hygroscopic material which can easily lead to injector
magnet excursion. Accordingly, the CsI samples were
deposited quickly on to a cathode of electrolytic
aluminium in a clean dry box. The cathode was then
baked in an oven at 100$^o$C for two days 
prior to the measurement.

\begin{figure}
\centerline{
\epsfig{file=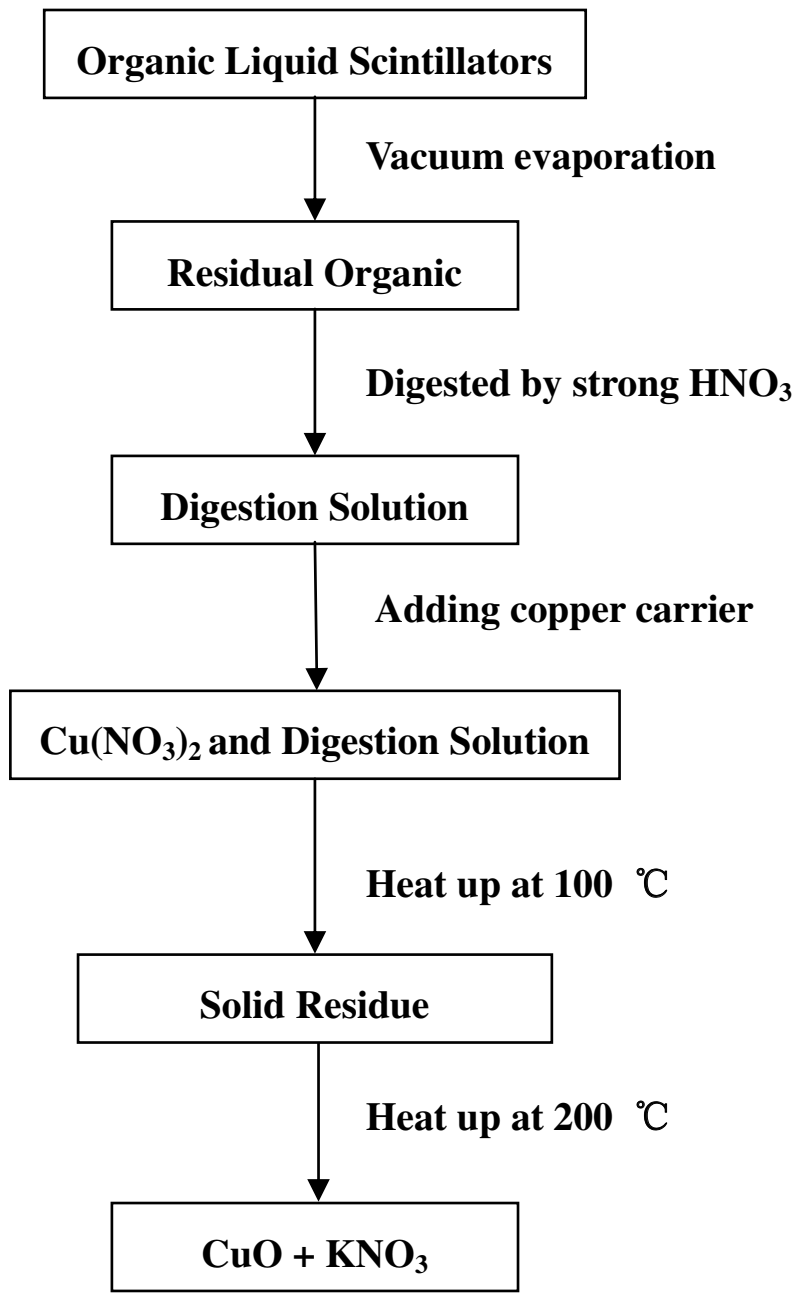,width=12cm}
}
\caption{
The flow chart of samples preparation for organic liquid scintillator.
}
\label{lsprocess}
\end{figure}

The LS pre-processing procedures were 
adapted from those of Refs.~\cite{munich}\&\cite{alabama}.
Different LS samples were prepared 
with different concentrations
of PPO dissolved in pure PC.
The procedures are summarized by the flow
chart of Figure~\ref{lsprocess}.
All steps were carried out in 
clean rooms better than the Class-100 grade.
Polypropylene bottles were etched in ultra-pure nitric acid and
ultra-pure hydrochloric acid over 5 days.
The bottles were then rinsed with 
ultra-pure water and then
pure PC, before left to be dried.
The LS samples were stored into these bottles
for further processing.

Quartz crucibles were etched 
with 65\% ultra-pure
hydrochloric acid and 65\% ultra-pure nitric acid 
over a period of one week
at sub-boiling  temperature.
Then they were
rinsed with 
65\% ultra-pure nitric acid 
and dried
at sub-boiling  temperature.
The LS samples were emptied to the crucibles
and evaporated under a vacuum of about 20~Torr.
The solid residual left behind was mostly
PPO powder, the least volatile component. 
The mass of solid residuals after evaporation
is typically less than 1\% of the LS samples.
Strong nitric acid was subsequently added
to extract the trace elements from the residuals.
The acid was heated to 80$^o$C on a hot plate.
At this temperature, the PPO samples 
fully dissolved into the acid,
forming a transparent yellow solution where the potassium
ions entered the acid phase.  
Ultra-pure copper\footnote{Supplier: Alfa Aesar}
with specified potassium concentrations
of less than 10~ppb was 
dissolved into the acid solution.
The $\rm{Cu (NO_3)_2}$ solution was evaporated slowly at 100$^o$C.
The residual solid was further baked at 200$^o$C for two hours.
The final substance was mostly CuO, while 
the trace amounts of potassium existed as
KNO$_3$.
These residual solid samples were used as target materials
and measured by AMS directly. 
Only a few volatile potassium compounds exist, 
such that good recovery of the potassium contaminations
from the original LS samples can be expected.

\subsection{Injector, Accelerator and Detector}

The $\kstable$  concentration in
the CsI powder and in the LS 
were measured with CIAE-AMS facility
depicted schematically in Figure~\ref{ciaeams}.
The tandem accelerator operated at a terminal
voltage of 8.0 MV.
A ``Multi-Cathode Source of Negative Ions
by Cesium Sputtering''
was used as the negative ion source.
Forty samples
were positioned on the target wheel at one time.
The wheel could be rotated
without affecting the vacuum  conditions
such that stable operating configurations were maintained
during measurements of a group of samples.
For normalizations, 
the intensities of the $^{127}$I$^-$ and
$\cu63 ^-$ relative to $\kstable ^-$ ions 
extracted with cesium sputter source 
were measured by the Low Energy Faraday Cup (LECup).

The extracted $\kstable ^-$ ions
were focused
by a trim einzel lens and a double
focusing 90$^o$ deflecting magnet where
the negative ion beams of interest
were momentum selected.
The ions were guided to
an aperture of 2~mm diameter
located at the entrance of the
pre-acceleration tube, where
they were accelerated up to
about 120 keV  kinetic energy.
The main accelerator operating at
a terminal voltage of 7.4~MV
further boost the ions to 
to 67~MeV. 
A carbon foil was attached
at the head of accelerator. The molecular
background was eliminated due to break-up of
molecular ions.

After passing through the accelerator
tank, ions with
charged state 8$^+$ were selected by a 90$^o$
double focusing analyzing magnet with a
mass energy product (M$\cdot$E/Z$^2$) at 200
to suppress the isotopic background.
A high-resolution electrostatic deflector
was placed at a branch beam line to
further reduce the isotopic
background and other undesired beam components.
The $\kstable$-ions which survived all selections
were detected by a gold-silicon surface barrier detector,
The $\kstable$ count rates as a function of time
were recorded for subsequent analysis.

\section{Measurements and Results}

\begin{figure}
\centerline{
\epsfig{file=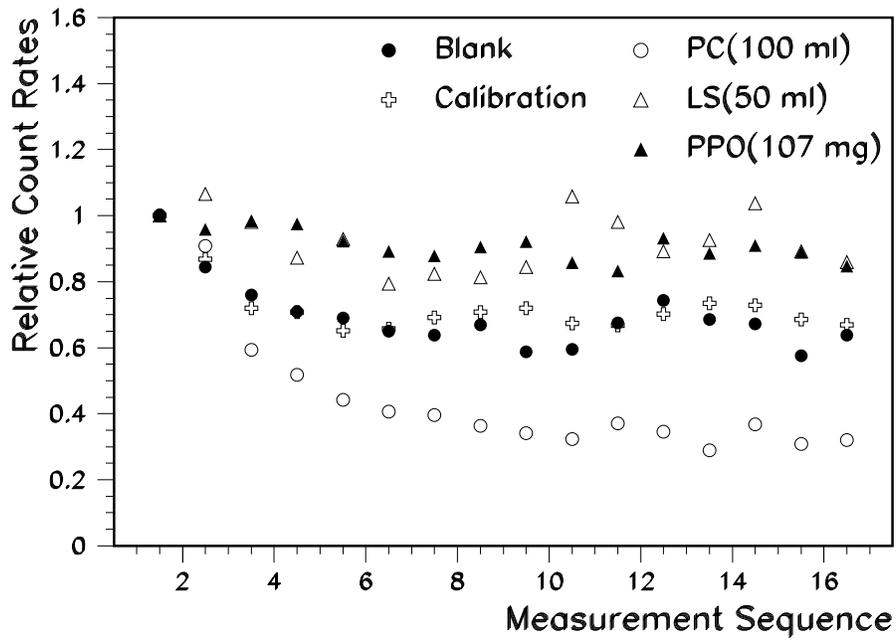,width=12cm}
}
\caption{
Time evolution of the
$\kstable$-counts for some selected
samples. Results after the
sixth measurements since the start
of data taking were adopted
for subsequent analysis.
}
\label{timeplot}
\end{figure}

Typical data taking of a certain sample
consisted of 16~sequential
measurements each lasting 100~s,
as depicted in Figure~\ref{timeplot} for some
selected samples.
Steady state was reached after several hundred seconds.
The quoted $\kstable$-levels
were derived from the means 
of those ten measurements from 600~s to 1600~s.
Typical statistical errors of individual measurement 
is less than 1\%, while the variance of the
ten measurements are of the order of 10\%.
The variances include systematic effects 
like unstable hardware conditions during data taking,
and are adopted as the
measurement uncertainties. 

\subsection{Blind Target}
\label{sect::beambkg}

In a typical AMS measurement, the samples
are placed inside a 1~mm diameter hole 
drilled out of a target structure made of
electrolytic aluminium. The negative
ions are then focused on to this hole.

\begin{figure}
\centerline{
\epsfig{file=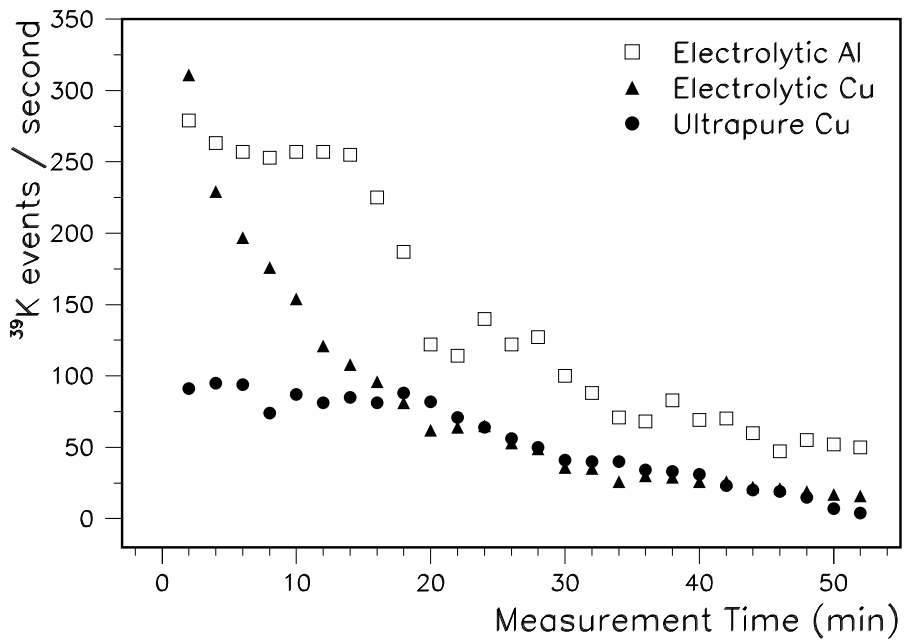,width=12cm}
}
\caption{
Time evolution of the
$\kstable$-counts for
the blind target runs.
}
\label{blanktarget}
\end{figure}

Measurements were performed with ``blind''
targets where the target structure
was replaced by a piece of solid metal.
This allowed measurements of the
intrinsic background due to ambient
contaminations and accelerator operation.
Three target materials were selected:
ultra-pure ($<$ppm) copper, electrolytic
copper and electrolytic aluminium. 
The variations of their
$\kstable$ background at the AMS detector
with time  are displayed in
Figure~\ref{blanktarget}.
The initial enhancement of background
are due to surface contaminations of the
target materials not relevant to subsequent
studies, while the steady-state
rates represents the upper bounds of 
the ion beam and
accelerator background.
The steady levels of the
order of 10~$\kstable$-events
per second are at least factors of 100's
less than the typical rates in
the measurements of various samples 
($\rm{> 1000 ~ s^{-1}}$),
such that the accelerator-related
background can be neglected.

\subsection{Inorganic CsI Powder}
\label{sect::csi}

The contaminations of $\kstable$ in CsI were directly measured 
by counting methods with the AMS detector, 
using KI as the reference sample.
For optimizations of the accelerator parameters on
$\kstable$ as well as the evaluations of 
the transmission efficiencies,
a similar isotope $^{37}$Cl was used. 
Negative $^{37}$Cl ions from AgCl
were emitted from the source 
and were accelerated by a terminal voltage
of 8.2~MeV. Charge states of $\rm{^{37} Cl ^{8+}}$ were selected.
The beam currents measured at the various stages are displayed
in Table~\ref{tab::clcurrent},
from which the transmission efficiencies can be evaluated.

\begin{table}
\begin{center}
\begin{tabular}{|l|c|cc|}
\hline
Measuring Cup & Beam Current  & 
\multicolumn{2}{c|}{\underline{Transmission Efficiency}} \\ 
& (nA) & This Stage & Cumulative \\ \hline \hline
Low Energy Cup & 900 & 1.00 & 1.00 \\
Image Cup & 350 & 0.39 & 0.39 \\
AMS Cup & 150 & 0.43 & 0.17 \\ \hline
\end{tabular}
\end{center}
\caption{
Beam currents and transmission efficiencies of 
the negative $^{37}$Cl ions.
}
\label{tab::clcurrent}
\end{table}

The two measurements on CsI and KI were normalized by
the LECup currents on the common $\rm{^{127}I^-}$ ions, 
denoted by $\rm{I_{LE} ( ^{127}I^- )}$.
The $\kstable$ contents in KI
were measured as beam currents in the Image Cup,
and compared to the $\kstable$-counts at the
AMS detector in the CsI run, taking into account
the transmission efficiency of 43\% between the
Image Cup and the AMS detector as shown in
Table~\ref{tab::clcurrent}.
The known relation of 
\begin{equation}
\label{eq::lecup}
\rm{
R_{LE}  ( ^{127}I^- ) 
~ = ~
6.25 \times 10^9 ~ \cdot
[ ~ I_{LE} ( ^{127}I^- ) /  nA  ~ ]
}
\end{equation}
was applied,
where $\rm{R_{LE}  ( ^{127}I^- ) }$ is the
$ \rm{^{127}I^-}$-counts
passing through the image cup
per second.  
The measured contamination levels of potassium are:
\begin{eqnarray}
\rm{ [ ~  \kstable  /  CsI  ~ ] }  & ~  = ~ & 
\rm{ ( 5.6 \pm  1.7 ) \times 10^{-7} ~ g/g ,
or ~ equivalently} \nonumber  \\
\rm{ [ ~ \k40  /  CsI ~ ] }  & ~ = ~ & 
\rm{ ( 7.0  \pm 2.1 ) \times 10^{-11} ~ g/g } ~.
\end{eqnarray}
This implies an activity $\k40$ decays in CsI 
of $\rm{1540 \pm 460~ kg^{-1} day^{-1}}$.
This result is consistent with 
and improves over that 
of an independent measurement on
the CsI powder 
by direct counting of the 1461~keV photons with
a high-purity germanium detector, 
where only an upper limit
\begin{equation}
\rm{  [ ~ \k40  /  CsI ~ ]  ~ < ~ 2 \times 10^{-10}  ~  g/g }
\end{equation}
could be derived.

As an illustrative comparison,
the intrinsic contaminations of $^{137}$Cs and the
U+Th series in CsI(Tl) crystals were recently
studied~\cite{csibkg}, where the levels of
\begin{eqnarray}
\rm{ [ ~ ^{137}Cs  /  CsI(Tl) ~ ] }  & ~ = ~ & 
\rm{ ( 1.7  \pm 0.3 ) \times 10^{-17} ~ g/g } ~ ~~ , \nonumber \\
\rm{ [ ~ ^{232}Th  /  CsI(Tl) ~ ] }  & ~ = ~ & 
\rm{ ( 2.23  \pm 0.06 ) \times 10^{-12} ~ g/g } ~ ~~ , \nonumber \\
\rm{ [ ~ ^{238}U  /  CsI(Tl) ~ ] }  & ~ = ~ & 
\rm{ ( 8.2  \pm 0.2 ) \times 10^{-13} ~ g/g } ~ ~~~ {\rm and } \nonumber \\
\rm{ [ ~ ^{235}U  /  CsI(Tl) ~ ] }  & ~ < ~ & 
\rm{ 4.9 \times 10^{-13} ~ g/g }  
\end{eqnarray}
were derived.
These results confirm the expectations that 
the potassium contaminations 
in most materials
are usually higher than those of
the other isotopes which also
exist in abundance in nature.

\subsection{Organic Solid and Liquid}

The results of the various runs with different
organic samples are summarized in
Table~\ref{lsresults}. 
As discussed in Section~\ref{sect::process}
and illustrated in Figure~\ref{lsprocess},
the chemical pre-processing of the samples
turned them into solid powder mixtures of
CuO+KNO$_3$.
Consequently, the AMS measurements of 
the $\kstable$
levels in the samples were all made
relative to $\cu63$.
The time evolution of the 16~measurements
for some selected samples are displayed in 
Figure~\ref{timeplot} as illustrations.
The raw data for each configuration include
the mean and variance of the $\kstable$-counts in 100~s,
s well as the $\cu63 ^-$ beam currents at
the LECup. 

All the 13 measurements listed in Table~\ref{lsresults}
were performed within 12~hours where the accelerator operation
has been stable.
The calibration sample served to establish a relationship
between $\kstable$-counts and $\cu63 ^-$ beam currents with
the known $\rm{ [ \kstable / Cu ]}$ contamination level. 
The good agreement of the cross-check samples with expected
results, as well as the internal consistencies
of the measured $\rm{[ \kstable / sample ]}$ 
levels among the
different categories of samples, demonstrate that the
experimental conditions were constant throughout the
measurements, and that a single calibration relation
can be applied consistently to all data.

The details of the measurements
are elaborated in the following sub-sections.

\begin{table}
\begin{center}
\begin{tabular}{|l|c|c|c|}
\hline
\bf{Samples} & 
\bf{$[$$\kstable$/Cu$]$} & \bf{$[$$\kstable$/Sample$]$} 
& \bf{$[$$\k40$/Sample$]$}   \\
&  \bf{(g/g)} & \bf{(g/g)} & \bf{(g/g)} \\ \hline \hline
\underline{Blank} &  & & \\
~~ No Sample & 
$\rm{ ( 1.6 \pm 0.1 ) \times 10^{-8}}$ & $-$  & $-$ \\ \hline
\underline{Calibration} & & & \\
~~ $[$$\kstable$/Cu$]$=1.5e-7~g/g & $-$ & $-$ & $-$ \\ \hline
\underline{Cross-Checks} & & & \\
~~ $[$$\kstable$/Cu$]$=1.5e-8~g/g & 
$\rm{ ( 1.7 \pm 0.2 ) \times 10^{-8}}$ & $-$ & $-$ \\
~~ $[$$\kstable$/Cu$]$=1.5e-9~g/g & 
$\rm{ ( 2.4 \pm 2.0 ) \times 10^{-9}}$ & $-$ & $-$ \\ \hline
\underline{PPO Powder} & & & \\
~~ mass=1.8~mg & 
$\rm{ ( 9.5 \pm 0.7 ) \times 10^{-8}}$ & 
$\rm{ ( 5.3 \pm 0.4 ) \times 10^{-7}}$ & 
$\rm{ ( 6.6 \pm 0.5 ) \times 10^{-11}}$  \\
~~ mass=19~mg & 
$\rm{ ( 9.8 \pm 1.0 ) \times 10^{-7}}$ & 
$\rm{ ( 5.2 \pm 0.5 ) \times 10^{-7}}$ & 
$\rm{ ( 6.5 \pm 0.7 ) \times 10^{-11}}$  \\
~~ mass=107~mg &  
$\rm{ ( 5.8 \pm 0.2 ) \times 10^{-6}}$ &
$\rm{ ( 5.4 \pm 0.2 ) \times 10^{-7}}$ & 
$\rm{ ( 6.8 \pm 0.3 ) \times 10^{-11}}$  \\  \hline
\underline{Pseudocumene} & & & \\
~~ volume=1~ml &  
$\rm{ ( -3 \pm 13 ) \times 10^{-10}}$ & 
$\rm{ < 2.2 \times 10^{-11}}$ & 
$\rm{ < 2.8 \times 10^{-15}}$ \\
~~ volume=5~ml 
&  $\rm{ ( 7 \pm 27 ) \times 10^{-10}}$  &
$\rm{ < 1.2 \times 10^{-11}}$ & 
$\rm{ < 1.5 \times 10^{-15}}$ \\
~~ volume=100~ml & 
$\rm{ ( 9.3 \pm 1.0 ) \times 10^{-8}}$  & 
$\rm{ ( 1.1 \pm 0.1 ) \times 10^{-11}}$ & 
$\rm{ ( 1.4 \pm 0.1 ) \times 10^{-15}}$  \\ \hline
\underline{Liquid Scintillator} & & & \\
~ (commercial) & & &  \\
~~ volume=1~ml &
$\rm{ ( 9.6 \pm 0.5 ) \times 10^{-8}}$ & 
$\rm{ ( 1.10 \pm 0.06 ) \times 10^{-9}}$ & 
$\rm{ ( 1.38 \pm 0.07 ) \times 10^{-13}}$  \\
~~ volume=5~ml & 
$\rm{ ( 4.4 \pm 0.2 ) \times 10^{-7}}$ & 
$\rm{ ( 1.01 \pm 0.04 ) \times 10^{-9}}$ & 
$\rm{ ( 1.27 \pm 0.06 ) \times 10^{-13}}$  \\
~~ volume=50~ml &  
$\rm{ ( 4.3 \pm 0.4 ) \times 10^{-6}}$ & 
$\rm{ ( 1.0 \pm 0.1 ) \times 10^{-9}}$ & 
$\rm{ ( 1.3 \pm 0.1 ) \times 10^{-13}}$  \\ \hline
\end{tabular}
\end{center}
\caption{
Summary of the results of the
various AMS measurements with
different configurations.
The two measurements with small volume of pseudocumene
did not give statistically significant results so that
only 90\% confidence level limits are quoted.
}
\label{lsresults}
\end{table}

\subsubsection{Blank (Sample-less) Measurement}
\label{sect::bkg}

A measurement was performed where the
exact procedures of Figure~\ref{lsprocess}
as described in Section~\ref{sect::process}
were followed $-$ except that there were no
initial samples. This allows the studies
of the intrinsic $\kstable$ background
due to the chemical processing procedures.

The AMS measurements gave
($\rm{ 73561 \pm 5565  }$) counts of $\kstable$ per
100~s at a $\cu63 ^-$ current of
778~nA at the LECup.
Using calibration factors derived in
Section~\ref{sect::calib}, this corresponds
to a background of 
\begin{equation}
\label{eq::bkg}
\rm{ 
 [ ~  \kstable / Cu ~ ] _{Bkg}  ~ = ~ ( 1.62 \pm 0.12 ) \times 10^{-8} ~ g/g ~~.
}
\end{equation}
After proper beam current normalizations are performed, 
this background  is subtracted off from the 
various AMS measurements, from which the
intrinsic $\kstable$ contaminations in the samples
can be derived.

As discussed in Section~\ref{sect::beambkg},
the beam-related background from the
blind target measurement is only
of the order of 10 $\kstable$ counts per second.
Accordingly, the measured $\kstable$
background of Eq.~\ref{eq::bkg}
is expected to be introduced 
pre-dominantly in the processing procedures.

\subsubsection{Reference Samples}
\label{sect::calib}

A series of potassium-copper standard samples
were prepared. 
The copper\footnote{Supplier: Alfa Aesar-A Johnson Matthey Company}
was of ultra-pure grade,
with purity level better than ppm and 
potassium contaminations less than 10~ppb.
Added to it were
65\% ultra-pure nitric acid and 99.99\% 
potassium nitrate (KNO$_3$).
The $\kstable$ levels were independently verified
by Spark Source Mass Spectrometry.
The exact processing procedures of Figure~\ref{lsprocess}.
were followed. 

The sample with highest potassium concentration, at 
$\rm{[ \kstable /Cu ] = 1.50 \times 10^{-7} ~ g/g}$,
was used for calibration purposes. 
The $\kstable$-counts at the AMS detector,
after subtraction of the intrinsic background of
Eq.~\ref{eq::bkg}, give rise to the 
$\kstable$ contamination level in the sample.
The number of copper atoms in the beam was derived
from Eq.~\ref{eq::lecup} using
the $\cu63 ^-$ beam current at the LECup 
($\rm{I_{LE} ( \cu63 ^- )}$)
as input.
The overall calibration factor is:
\begin{equation}
\label{eq::conversion}
\rm{
[ \kstable / Cu ] ~ = ~ 
1.5 \times 10^{-7} \cdot
\frac{R_{AMS} ( \kstable )}{0.92 \times 10^6} \cdot
\frac{I_{LE} ( \cu63 ^- )}{1000~nA}  
~~ g/g  ~~,
}
\end{equation}
where 
$\rm{R_{AMS} ( \kstable ) }$ is the $\kstable$-counts measured
in the AMS detector in 100~s.
This relation was applied in all the other measurements
listed in Table~\ref{lsresults}.

Two additional samples with 10 and 100 times lower
potassium concentrations were prepared 
to cross-check the procedures.
The AMS measurement results, shown
in Table~\ref{lsresults}, 
were derived
after applying this conversion factor 
and subtracting the normalized background. 
They are in excellent
agreement with the expected values.

The $\kstable$-levels in
$\rm{ [ \kstable / sample ]}$ units for
the subsequent samples can be derived through
\begin{equation}
\label{eq::massscaling}
\rm{
 [ ~ \kstable / sample ~ ] ~ = ~ 
 [ ~ \kstable / Cu ~ ] \cdot
[ ~  \frac{ M (Cu) }{ M (sample) } ~ ]  
}
\end{equation}  
where M(Cu) and M(sample) are the
controlled and known masses of
copper and the sample, respectively.
It can be seen that given a sample
with definite $\rm{ [  \kstable / sample ] }$,
the uncertainties in the measurement of
$\rm{[  \kstable / Cu ]}$ can be reduced by
using larger sample mass for the same
amount of copper carrier introduced.
The final results in $\rm{ [ \k40 / sample ] }$
follow simply from taking ratios on the
isotopic abundance between $\k40$ and $\kstable$.

\subsubsection{PPO Concentration Scan}

\begin{table}
\begin{center}
\begin{tabular}{|l|c|c|c|}
\hline
Samples & $\alpha$ & $[$$\kstable$/sample$]$ ( g/g ) 
& $\epsilon$ \\ \hline \hline
PPO Solid & 100\% &
$\rm{Y_{PPO} = ( 3.1 \pm 0.4 ) \times 10^{-8}}$ & $-$ \\
Pure PC & 0.0\%  &
$\rm{Y_{0} = ( 1.5 \pm 0.2 ) \times 10^{-11} }$ & $-$ \\
PC+PPO & 0.01\%  &
$\rm{Y_{\alpha} = ( 1.8  \pm 0.3 ) \times 10^{-11}}$ & $1.0 \pm 1.3$ \\
PC+PPO & 0.1\%  &
$\rm{Y_{\alpha} = ( 4.4 \pm 0.8 ) \times 10^{-11} }$ & $1.0 \pm 0.3$ \\
PC+PPO & 1.0\%  &
$\rm{Y_{\alpha} = ( 3.2 \pm 0.1 ) \times 10^{-10} }$ & $1.0 \pm 0.1$ \\ \hline
\end{tabular}
\end{center}
\caption{
Results from AMS measurements with varying PPO concentration in
PC.
}
\label{tab::pposcan}
\end{table}

The relation established in Eq.~\ref{eq::conversion}
is applicable for calibrations of 
AMS measurements in solid samples
following the processing procedures of 
Figure~\ref{lsprocess}.
For organic liquid samples, an extra factor 
$-$ the survival efficiency of potassium 
at the solid residuals 
after vacuum evaporation ($\epsilon$) $-$
has to be evaluated.
This was studied by measuring the $\kstable$-yields
in solid PPO ($\rm{Y_{PPO}}$) and comparing it 
to those samples with PPO mixed with PC.
The efficiency $\epsilon$ is given by
\begin{equation}
\rm{
\epsilon ~ = ~
\frac{ Y_{\alpha} - Y_0 }{ \alpha \cdot Y_{PPO} }
}
\end{equation}
where $\rm{Y_{PPO} ~ , ~ Y_0 ~ and ~ Y_{\alpha}}$ are, 
respectively, the 
$\kstable$-levels in PPO solid, pure PC liquid without PPO, 
and PC mixed with PPO in known fractions $\alpha$. 
The results are summarized in Table~\ref{tab::pposcan}.
The most accurate measurement is one with a high concentration
of PPO ($\alpha$=1\%), giving $\epsilon = ( 1.0 \pm 0.1 ) $. 
This result demonstrates that most potassium 
in the PPO+PC solution remained in the solid residuals
after evaporation. Measurements with smaller PPO concentrations
were still consistent with complete potassium extraction,
but with larger uncertainties due to limited statistics above
the background.

\subsubsection{Liquid Scintillator, Solute and Solvent}

Samples of different masses of
pure PPO powder, pure PC liquid, as well
as their mixtures as liquid scintillators
were measured, following the proper
procedures
of calibrations and background subtraction
discussed above.
The results are summarized
in Table~\ref{lsresults}.
Consistent results were obtained
for the two decades variations in
sampling mass, in both 
$\rm{ [ \kstable / Cu ]}$ 
and $\rm{ [ \k40 / sample ]}$.

The liquid scintillator sample was commercially mixed.
The measured $\k40$ contamination was
\begin{equation}
\rm{
[ ~ \k40 / LS ~ ] ~ = ~ ( 1.3 \pm 0.1 ) \times 10^{-13} ~ g/g 
} 
\end{equation}
implying equivalently a $\k40$-activity of
$\rm{ \sim 3 ~ kg ^{-1} day^{-1}}$.
Such commercial LS sample is not yet good enough
for the low background experiments. 
Further purifications are therefore necessary. 
The source of contaminations come primarily from 
the PPO solute, such that it would be more 
effective to focus the purification program on it.
The best measurement is, expectedly,
one on pure PC at the largest sampling mass, where 
\begin{equation}
\label{eq::k40pc}
\rm{
[ ~ \k40 / PC ~ ] ~ = ~ ( 1.4 \pm 0.1 ) \times 10^{-15} ~ g/g 
} ~.
\end{equation}
The expected $\k40$-decay rate is 
$\rm{ ( 0.024 \pm 0.003 ) ~ kg ^{-1} day^{-1}}$.
These characteristic features should be valid in general $-$
that the contaminations in organic liquid scintillator
come primarily from the dye solutes while the
liquid solvent are orders of magnitude cleaner.
Another two measurements of PC with much smaller
sampling masses did not give rise to statistically
significant excess of the $\rm{ [ \kstable / Cu ]}$ 
ratio, such
that only 90\% confidence level upper limits are
derived~\cite{pdg06} in Table~\ref{lsresults}. 
The limits are consistent with the
measured results of Eq.~\ref{eq::k40pc}.

It is instructive to compare the
results of Tables~\ref{lsresults}\&\ref{tab::pposcan},
which are from the measurements on
two different batches of PPO and PC samples
from the same supplier.
The $\kstable$ contaminations of the
two batches of PPO are, respectively, 
$\rm{Y_{PPO} = ( 5.4 \pm 0.2 ) \times 10^{-7} ~g/g }$
and $\rm{( 3.1 \pm 0.4 ) \times 10^{-8} ~g/g }$,
while the those for the PC are
$\rm{Y_{0} = ( 1.1 \pm 0.1 ) \times 10^{-11} ~g/g }$
and $\rm{( 1.5 \pm 0.2 ) \times 10^{-11} ~g/g }$.
The differences indicate the range of variations
of potassium levels in different production 
batches among the same materials.

\subsection{Limiting Sensitivities}

An important parameter to characterize the performance of
this measurement technique is the limiting sensitivity.
It represents the upper limits that can be set for samples
where no excess of $\kstable$-counts above background
are observed $-$ or alternatively, the minimal 
contamination levels that will produce a positive
measurement. 

The measurements of the CsI samples do not
involve chemical pre-processing. Therefore, the
sensitivity is limited by the ion-beam related
background measured in the ``blind target''
run discussed in Section~\ref{sect::beambkg}.
The $\kstable$
background count rates of $<$50~per second
at the AMS detector
can be translated to limiting sensitivities
following the same normalization steps of
Section~\ref{sect::csi}:
\begin{eqnarray}
\rm{ [ ~  \kstable  /  CsI ~ ] _{limit} }
 & ~ \sim  ~ & 
\rm{ 6 \times 10^{-9} ~ g/g , ~~
or ~ equivalently} \nonumber \\
\rm{ [ ~  \k40  /  CsI ~ ] _{limit} }
& ~ \sim ~ & 
\rm{ 8 \times 10^{-13} ~ g/g } ~.
\label{eq::limitcsi}
\end{eqnarray}

For organic liquid,
potassium introduced during the pre-processing
procedures is the dominant background.
The limiting sensitivities
of $\rm{[ \kstable / Cu ]}$ 
can be derived from the uncertainties of
the background blank measurement in Eq.~\ref{eq::bkg},
giving
\begin{equation}
\rm{
[ ~  \kstable / Cu ~ ] _{limit} ~ \sim ~  2 \times 10^{-9} ~ g/g  ~ .
}
\end{equation}
To convert into [$\kstable$/(Org.Liq.)] units, 
the relation of Eq.~\ref{eq::conversion}
is used.
The ratio of [M(sample)/M(Cu)] is taken from the
measurement with the largest liquid sample volume (100~ml)
where the extraction of the contaminants was
successfully demonstrated. 
The sensitivities are
\begin{eqnarray}
\rm{ [ ~  \kstable  /  ( Org. Liq. ) ~ ] _{limit} }
 & ~ \sim ~ & 
\rm{ 2 \times 10^{-13} ~ g/g , ~~
or ~ equivalently} \nonumber \\
\rm{ [ ~  \k40  / ( Org. Liq. ) ~ ] _{limit} }
& ~ \sim ~ & 
\rm{ 3 \times 10^{-17} ~ g/g } ~.
\label{eq::limitorgliq}
\end{eqnarray}

The improvement of 
sensitivities 
in Eqs.~\ref{eq::limitorgliq}\&\ref{eq::limitcsi}
of organic liquid over solid CsI powder
follows from the
[M(Cu)/M(sample)] ratio in
Eq.~\ref{eq::massscaling}.
At a typical M(Cu)$\sim$10~mg  
and the largest liquid extraction volume of
100~ml [that is, M(sample)$\sim$100~g], 
an enhancement factor of
$10^{-4}$ is readily attained
in the evaporation and extraction processes.

\section{Summary and Outlook}

In this article, we report 
the first measurement of ultra-low
potassium contaminations in an inorganic salt (CsI)
and in an organic scintillator liquid (LS=PC+PPO)
using accelerator mass spectrometry.
Comparable sensitivities to the neutron activation
techniques were achieved. 
Potassium were positively identified in 
samples of commercially available CsI and LS,
at the contamination levels of 
$7 \times 10^{-11} ~ {\rm g/g}$ and
$1.3 \times 10^{-13} ~ {\rm g/g}$, respectively. 
The radioactivity in the LS is mainly due to
the dye solutes.
The limiting sensitivities of the techniques
are of the order
of $10^{-13}$~g/g of $\k40$ in solid powder and
of $10^{-17}$~g/g of $\k40$ in an organic
liquid.

In the case of organic liquid, the sensitivities 
are limited by ambient potassium impurities
introduced during
the chemical processing of the liquid
depicted in Figure~\ref{lsprocess}.
Further improvement can be made along
these directions.
The beam-related background is small in
comparison. 
A boost in sensitivities can also
be made with the use of larger initial
liquid volume for evaporation. 
The largest sample demonstrated in the present studies
is 100~ml.  
Extensions to treatment of liter-scale sample 
are expected to be technically
similar, such that a limiting sensitivity of
$\sim 10^{-18} ~{\rm g/g}$ on liquid scintillator
using AMS should be possible.

This work demonstrate the feasibilities of 
measuring ultra-low impurities of 
naturally-occurring yet unstable
isotopes with AMS. 
Accordingly, other radio-isotopes which
are problematic to low-background experiments,
such as the $^{238}$U and $^{238}$Th series
as well as $^{87}$Rb, $^{113}$Cd,
$^{115}$In, $^{138}$La and $^{176}$Lu, 
can also be measured by the AMS techniques.
The goals of our future efforts would be to
devise schemes to measure some of these isotopes.
Extensions to Rb will be relatively straight-forward,
while the measurements of the $^{238}$U and $^{238}$Th series
will involve upgrades of the present accelerator systems.

\section{Acknowledgments}

The authors are grateful to
M.~Lin, S.H.~Li, Y.D.~Chen, A.L.~Li, J.~Li,
K.X.~Liu, B.F.~Ni, H.Q.~Tang, W.Z.~Tian and Z.Y.~Zhou
for helpful discussions, as well as to
Q.B.~You, Y.B.~Bao, Y.M.~Hu
and the technical staff at CIAE
for smooth accelerator operation. 
This work is supported by 
contract 10375099 from the National Natural
Science Foundation, China.

\newpage
                                                                                
\end{document}